\documentclass[acus]{JAC2000}   
\usepackage{graphicx}     

\newcommand{\be}{\begin{equation}}
\newcommand{\ee}{\end{equation}}
\newcommand{\bea}{\begin{eqnarray}}
\newcommand{\eea}{\end{eqnarray}}

\newcommand{\bc}{\begin{center}}
\newcommand{\ec}{\end{center}}

\begin{document}
\title{SUPPRESSION OF BEAM BREAKUP INSTABILITY IN A LONG TRAIN BY INTRODUCING
ENERGY SPREAD BETWEEN THE BUNCHES\thanks{Work supported by
DOE contract DE-AC03-76SF00515.}}
\author{G. V. Stupakov, SLAC, Stanford, CA 94309, USA}

\maketitle
\begin{abstract}

Interaction between the bunches in the NLC main linac via long range
wakefields can cause a beam breakup instability. Although the
magnitude of the long range wakefields for an ideal NLC structure is
below the instability threshold, the wake increases when structure
manufacturing errors are taken into account. In case of large errors,
the developing instability can result in the projected emittance
dilution of the bunch train. To suppress the instability, we propose
to introduce an energy spread between the bunches, similar to the BNS
energy spread for damping of the beam breakup within a single bunch
\cite{bns}. Based on simple estimates, we show that the energy spread
of order of 1-2\% should be enough for suppression of the
instability. The results of computer simulations with the simulation
code LIAR confirm theoretical estimates and indicate that both the
tolerances for structure misalignments and the incoming beam jitter
can be considerably loosened by introducing the energy spread within
the beam.

\end{abstract}
\section{Introduction}

Interaction between the bunches in the NLC linac via the long range
wakefields can cause a beam breakup instability \cite{thompson90r},
similar to the beam breakup within a single bunch that is caused by
short range wakes \cite{chao80ry}. This instability, if severe,
imposes tight tolerances on the beam injection errors as well as
structure misalignments in the lattice.

\begin{figure}[ht]
        \begin{center}
        \includegraphics[scale=0.7]{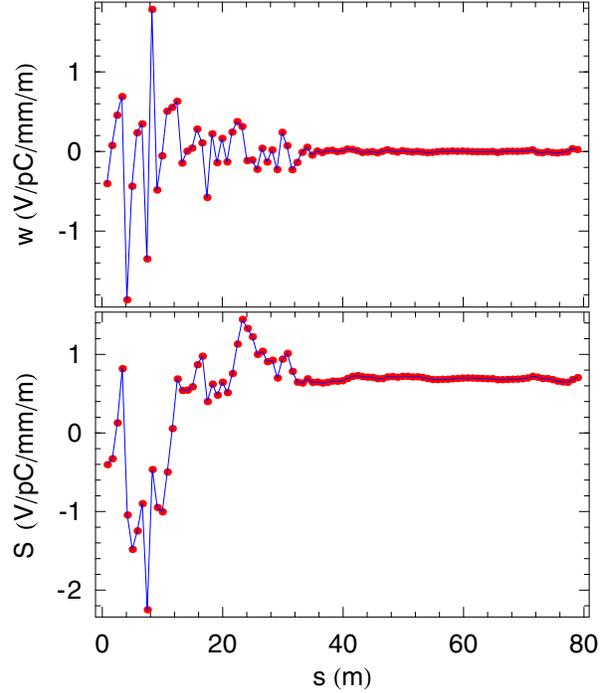}
        \end{center}
        \caption
        {
        Transverse wake $w$ and the corresponding sum wake $S$
        for NLC linac with 3 MHz rms frequency error in the dipole
        modes. Each point corresponds to a single bunch. The
        total number of bunches in the train is 95 and the spacing
        between the bunches is 83 cm.
        \label{wake}
        }
        \end{figure}

An example of the the long-range wake for the NLC structures is shown
in Fig. \ref{wake}. This wake was calculated assuming  a random
frequency error of the dipole mode in the structure with the rms
spread of 3 MHz \cite{roger}. Note that in contrast to the
short-range wakefield, that can usually  be approximated by a linear
function, the long-range wake is a complicated oscillating function
of the bunch position.

One can try to estimate the projected normalized emittance growth
$\Delta \epsilon$ of a train of bunches caused by randomly misaligned
structures in the linac using the result of Ref. \cite{bane_et_al}
for the expectation value of $\Delta \epsilon$
     \be \label{emit_growth}
     \langle{\Delta \epsilon}\rangle
     =\delta y^2 r_{e}^{2}N^{2}\bar{\beta}_{0}N_{\mathrm s}
     L_{\mathrm s}^{2}
     \langle \Delta S^{2}\rangle
     \frac{1-(\gamma_{0}/\gamma_{f})^{1/2}}
     {\gamma_{0}^{1/2}\gamma_{f}^{3/2}}\,,
     \ee
where $\delta y$ is the rms structures offset in the linac, $r_e$ is
the classical electron radius, $N$ is the number of particles in the
bunch, $\bar{\beta}_0$ is the average value of the beta function at
the beginning of the linac, $N_{\mathrm s}$ is the number of
structures in the linac, $L_{\mathrm s}$ is the length of the
structure, $\gamma_{0}$ and $\gamma_{f}$ are the initial and final
relativistic factors of the beam, and $S$ is the sum wake. The $k$-th
component of $S$ is defined  as a sum of the transverse wakes $w_i$
generated by all bunches preceding the bunch number $k$ (with
$S_1=w_1=0$),
     \be \label{Sk}
     S_{k}=\sum_{i=1}^{k}w_{i}\,.
     \ee
The quantity $\Delta S$ is the difference between $S$ and the average
value $\langle S\rangle$, $\Delta S_k=S_k-\langle S\rangle$, with
     \be \label{deltaS}
     \langle S\rangle={1\over N_{b}}\sum_{k=1}^{N_{b}}S_{k}\,,
     \qquad
     \langle \Delta S^2\rangle
     ={1\over N_{b}}\sum_{k=1}^{N_{b}}\Delta S_{k}^2\,,
     \ee
where  $N_{b}$ is the number of bunches. Eq. (\ref{emit_growth}) is
derived assuming a lattice with the beta function smoothly increasing
along the linac as $\bar{\beta} \sim E^{1/2}$. Note that Eq.
(\ref{emit_growth}) is only valid if there is no beam break up
instability, and hence gives a minimal emittance growth for given
misalignment.

For the wake shown in Fig.~\ref{wake}, $\langle \Delta
S^{2}\rangle^{1/2}=0.59$ V/pC/m/mm. Using the nominal NLC
parameters: $N=1.1\cdot 10^{10}$, $\bar{\beta} = 7$ m, $N_s = 4969$,
$L_s = 1.8$ m, $E_0 = 10$ GeV, $E_f = 500$ GeV, and assuming $\delta
y = 10\,\,\mu$m one finds $\langle{\Delta \epsilon}\rangle = 2.9\cdot
10^{-10}$ m, which is about 1\% of the nominal beam emittance at the
beginning of the linac, $\epsilon = 3\cdot 10^{-8}$ m.

However, computer simulations for this case with the wake shown in
Fig.~\ref{wake} demonstrate the projected emittance growth in the
train of order of 50\%, (see Fig.~\ref{emit_growth}), that is much
larger than the above estimate. The reason for such large emittance
dilution is the development of the beam breakup instability due to
the long-range wakefields in the bunch train reflected in
quasi-exponential growth in Fig.~\ref{emit_growth}.
        \begin{figure}[ht]
        \begin{center}
        \includegraphics[scale=0.55]{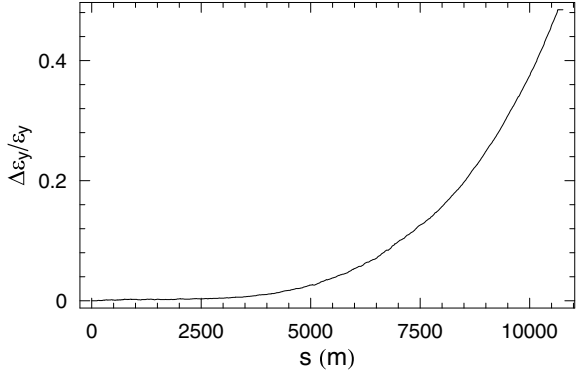}
        \end{center}
        \caption
        {
        Relative emittance growth along the linac in the train of
        bunches, for 10 $\mu$m rms structure misalignments,
        when all bunches have the same energy. The bunches
        are simulated as single particles, so that there is no
        emittance growth within the bunch.
        \label{emit_growth}
        }
        \end{figure}
To suppress the instability, one can try to introduce an energy
spread in the train of bunches \cite{stupakov}, similar to the energy
spread within a bunch that is routinely use for the BNS damping of a
single bunch beam breakup.

\section{Estimate of Required Energy Spread}

To obtain a rough estimate of the energy spread required to suppress
the long range beam break up instability, we use the autophasing
condition for the BNS damping in the two-particle model for a FODO
lattice \cite{chao93,ZDR}
    \be \label{BNS}
    \delta = \frac{Ne^2wl_{cell}^2}{24 E}
    \left(1+\frac{3}{2}\mathrm{ctn}^2\frac{\mu}{2}\right),
    \ee
where $\delta$ is the relative energy difference between the
particles, $N$ is the number of particles in the macroparticle, $w$
is the transverse wake, $E$ is the beam energy, $\mu$ is the betatron
phase advance per cell in the FODO lattice, and $l_{cell}$ is the
cell length. The value of $\delta$ given by Eq. (\ref{BNS}) would be
enough for suppression of the beam break up instability between the
two macroparticles with the interaction characterized by the wake
$w$.

\begin{figure}[ht]
        \begin{center}
        \includegraphics[scale=0.65]{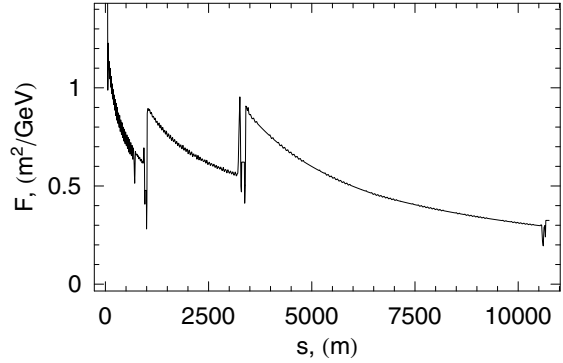}
        \end{center}
        \caption
        {
        Function $F(s)$ for the NLC lattice.
        \label{F}
        }
        \end{figure}

To apply Eq. (\ref{BNS}) for a train of bunches, we use the rms value
of the wake shown in Fig.~\ref{wake}, $w_{\mathrm{rms}}= 0.36$
V/pC/m/mm. The quantity $F\equiv{l_{cell}^2}\left(1+\frac{3}{2}
\mathrm{ctn}^2\frac{\mu}{2}\right)/(24E)$ from Eq. (\ref{BNS}) was
calculated for the NLC lattice as a function of distance $s$, see
Fig. \ref{F}. For the estimate, we use the average value of
$F_{av}=0.54$ m$^2$/GeV.
Finally, we need to relate the quantity $\delta$ to the energy spread
in the train. Since Eq. (\ref{BNS}) was derived for two
macroparticles, one can expect \emph{that $\delta$ refers to the
energy difference between the adjacent bunches in the train}. Hence,
the required  energy spread in the train for suppression of the
instability is equal to $\delta$ multiplied by the total number of
bunches in the train, $\delta_\mathrm{train} \approx
N_{\mathrm{b}}\delta$. We can now estimate $\delta_\mathrm{train}$ as
$N_{\mathrm{b}}Ne^2w_{\mathrm{rms}}F_{av}$, which gives
    \be
    \delta_\mathrm{train} \approx 0.03.
    \ee
This should be compared with the energy spread of the order of 1\%
within the bunch introduced for the BNS damping of the short-range
beam breakup instability.

A more detailed theory of the beam breakup instability in a train
with the energy chirp can be found in Ref. \cite{bohn}.

\section{Computer Simulations}

We carried out computer simulations using LIAR code \cite{liar} with
variable energy spread in the train for the wake shown in Fig.
\ref{wake}. In these simulations, the bunches were treated as
macroparticles  to suppress the effect of short range wakefields.
        \begin{figure}[ht]
        \begin{center}
        \includegraphics[scale=0.7]{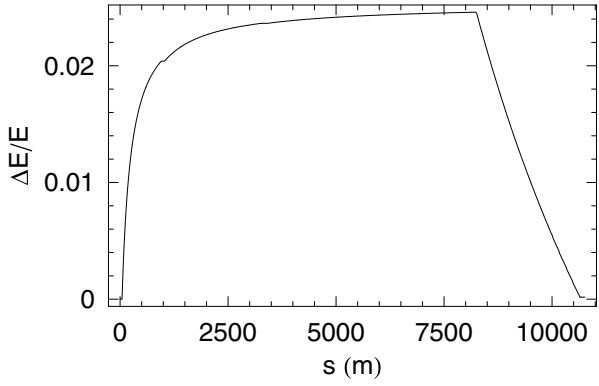}
        \end{center}
        \caption
        {Variation of the relative energy spread $\Delta E/E$
        along the linac.
        \label{slope}
        }
        \end{figure}
We used two types of energy profiles in the train. In the first case,
the energy in the train varied linearly so that $\Delta E_k = -e(s)
k$, where $k$ is the bunch number in the train, and $e(s)$ is the
energy slope.
The rms energy spread in the train $\Delta E$ in this case is $\Delta
E(s)= |e(s)|N_{\mathrm{b}}/\sqrt{12}$. An example of the profile of
$\Delta E(s)$ along the linac is shown in Fig.~\ref{slope}
--- the energy spread was generated at the beginning of the linac,
and taken out at the end, so that the final energy spread was close
to zero. We also used an exponential energy profile in the train
$\Delta E_k = e(s)[1-\exp(-3k/N_\mathrm{b})]$ with the same
functional form of the rms $\Delta E (s)$ as shown in
Fig.~\ref{slope}.
        \begin{figure}[ht]
        \begin{center}
        \includegraphics[scale=0.7]
           {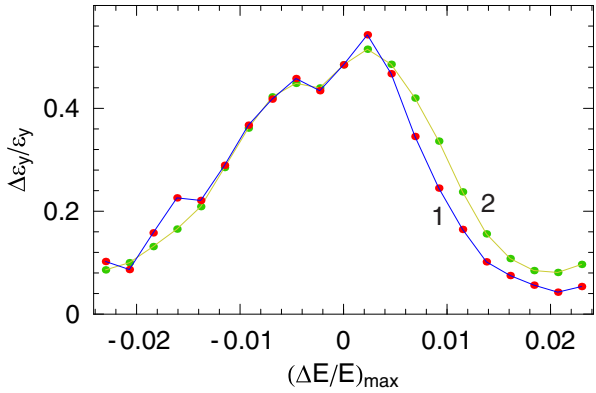}
        \end{center}
        \caption
        {
        Relative emittance increase in the train of bunches
        as a function of the maximum rms energy spread in the train.
        Curve 1 corresponds to a linear
        variation of the energy, and curve 2 --- to exponential
        function $\Delta E_k$.
        \label{emit_growth_vs_energy}
        }
        \end{figure}

In the first set of simulations, all  structures were randomly
misaligned with rms value of 10 $\mu$m. The  energy spread in the
train varied by scaling the profile shown in Fig.~\ref{slope} and the
final effective emittance of the train was calculated as a function
of the maximum rms energy spread  $(\Delta E/E)_\mathrm{max}$
(approximately at $s\approx 8000$ m). The result is shown in Fig.
\ref{emit_growth_vs_energy}. Positive values of $(\Delta
E/E)_\mathrm{max}$ correspond to the BNS-like energy profile, when
the energy decreases toward the tail of the bunch and negative values
of  $(\Delta E/E)_\mathrm{max}$ correspond to the opposite slope of
the energy profile. As we see, the positive values of $(\Delta
E/E)_\mathrm{max}$ are more effective in suppression of the emittance
growth.

\begin{figure}[ht]
        \begin{center}
        \includegraphics[scale=0.7]
          {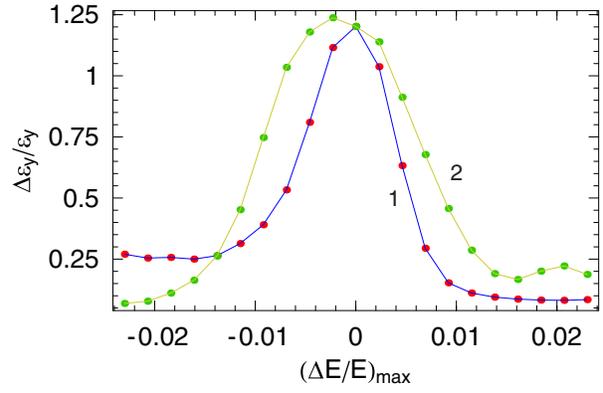}
        \end{center}
        \caption
        {
        Relative emittance increase as a function of the  energy spread for
        initial train offset of 1 $\mu$m. Curve 1 corresponds to a linear
        variation of the energy, and curve 2 --- to exponential
        function $\Delta E_k$.
        \label{emit_growth_vs_energy_init_offset}
        }
        \end{figure}

In the second set of simulations, the train was initially offset by 1
$\mu$m (with all structures and quadrupoles perfectly aligned). The
resulting emittance growth of the beam as a function of $(\Delta E/E)_\mathrm{max}$ is
shown in Fig. \ref{emit_growth_vs_energy_init_offset}. Again, we see
that the energy spread of the order of 1\% results in much smaller
emittance growth then for the bunches with the same energy.

\section{Discussion}

An energy spread between the bunches in the NLC is naturally
generated in the linac by the beam loading effect. A special
compensation scheme in the present NLC design will correct the energy
spread to a minimal value. With a slight modification of the
compensation scheme, it should also able to introduce a small
controllable energy spread required for the long-range BNS damping,
as studied in this paper. Another option of generating the energy
spread between the bunches is using RFQ magnets.

\section{Acknowledgements}

The author thanks T. Raubenheimer for useful discussions.

\end{document}